\documentstyle[11pt,moriond,epsfig]{article}

\def\Journal#1#2#3#4{{#1} {\bf #2}, #3 (#4)}

\def\NPB{{\em Nucl. Phys.} B}

\def\PLB{{\em Phys. Lett.}  B}

\def\PRL{\em Phys. Rev. Lett.}

\def\PRD{{\em Phys. Rev.} D}

\def\bqn{\begin{equation}}

\def\eqn{\end{equation}}

\def\bea{\begin{eqnarray}}

\def\eea{\end{eqnarray}}

\begin{document}

\vspace*{4cm}

\title{Time varying gauge couplings and di-nucleon states}

\author{Malcolm Fairbairn}

\address{Service de Physique Th\'eorique, CP225,\\Universit\'e Libre de Bruxelles, B-1050 Brussels, Belgium}

\maketitle\abstracts{In this talk we discuss the effect upon the di-proton, di-neutron and deuteron of a time variation of the QCD gauge coupling.  We describe how a time evolution of the size of extra dimensions can give rise to such a variation and show how this is related to the recent results on possible time variation of the electromagnetic fine structure constant.  Based on work in collaboration with Tom Dent.}

\section{Introduction}

Webb et al have observed the absorption of light from quasars by intervening
intergalactic clouds of matter at high redshifts \cite{results}.  By
studying the relative position of absorption lines in these clouds, they have
found that there appears to be a variation in the electromagnetic fine
structure constant $\alpha_{em}$ at high redshift.  

Such a variation is
naturally accommodated in higher dimensional theories such as string or
M-theory since in these models the value of the gauge couplings are set by
the size of the compactified directions.  Cosmological evolution of the higher dimensions would then lead to time variation of the gauge coupling of the grand unified theory.  One would expect this to lead to a variation of all of the gauge couplings
in the standard model including the QCD coupling.  

The renormalisation group running of the QCD
coupling constant $\alpha_3$ means that the accompanying fractional variation in the characteristic energy scale of hadronic physics $\Lambda_{QCD}$ would be much larger than that for $\alpha_{em}$.  It is natural therefore to investigate as to whether or not the observed variation is consistent with the assumptions of gauge coupling unification and physics in the hadronic sector.

After briefly discussing the results, we talk more about the relationship between the size of higher dimensions and the gauge couplings in the 4D theory in one particular model.
Next, we describe how variations in $\alpha_{em}$ lead to variations in
$\Lambda_{QCD}$ and the implications of this for the binding of the two
nucleon states, the deuteron, di-neutron and di-proton.  We then show that this leads to constraints on the variation of $\alpha_{em}$ from the variation of the binding energy of the deuteron during nucleosynthesis.

\section{The observations}
As we have said, Webb et al\cite{results} have measured $\alpha_{em}$ at high redshifts by observing the absorption lines in the light from high redshift quasars.  This absorption occurs in clouds of intervening intergalactic dust.  When trying to understand the origin of a time variation in $\alpha_{em}$, it is useful to know a tiny part of the atomic physics that allow one to probe such a variation.
  
Upon solution of Schrodinger's equation in atomic physics, the $n$th electron energy level of a given atom is proportional to
\begin{equation}
E\propto-\frac{me^4}{2\hbar^2}\frac{Z^2}{n^2}
\end{equation}
where $m$ is the electron mass, $e$ is it's charge and $Z$ is the number of
protons in the nucleus.  The relativistic correction to this energy level has
the following proportionality \cite{dzuba}
\begin{equation}\label{eq:relcor}
\Delta \propto
\frac{me^4}{2\hbar^2}\frac{Z^2}{n^2}\frac{(Z\alpha_{em})^2}{n}.
\end{equation}
Comparison of the previous two equations is the key to understanding why
observation of spectral lines in high redshift objects is an
extremely sensitive probe of the variation of the gauge couplings.  Typically,
if one was trying to determine the energy of a particular spectral line in a
high redshift object, there would be many systematic errors due to
uncertainties in one's knowledge of the redshift or proper motion of the
object.  However, these systematic errors fall out of the analysis if one
deduces $\alpha_{em}$ by comparing a number of lines within the same cloud.
It is also interesting to note that such measurements are not sensitive to a time variation in the mass of the electron, since this would simply act like an apparent change in the redshift of the system.

It is then possible to compare the observed lines to those predicted
numerically to see if the best fit is with $\alpha_{em}$ as we know it today,
or with a different value of $\alpha_{em}$.  The analysis of Webb et al\cite{results}
shows that for redshifts of approximately $z\sim 1-3$ the observed spectral lines are
statistically significantly better fit by the numerical models if one assumes
that $\alpha_{em}$ was smaller by about $1$ part in $10^5$ at those times.  

It is therefore interesting to consider the effect of this variation assuming that the gauge fields are part of some larger GUT symmetry and that variation of one gauge coupling leads to variation of all three gauge couplings.

\section{GUT theories and higher dimensions}
The observed unification of the gauge couplings at $10^{16}$ GeV in the supersymmetric (SUSY) version of the standard model strongly\cite{ross} suggests that both QCD and the electroweak theory are parts of a larger gauge symmetry broken at high energy.  In this section we show how the coupling of the GUT gauge field is related to the size of extra dimensions in one particular model, namely Heterotic M-theory.  We then report how this all relates to $\alpha_{em}$ and the hadronic sector following the analysis of our paper\cite{ourpaper}.

\subsection{Some (M)otivation}
We know that in order to have general relativity and some unified gauge theory with field strength $F$
containing the standard model, the Lagrangian of the effective four dimensional theory obtained
by compactification of some particular string/M-theory must look like
\begin{equation}
L_{eff}=-\int d^{4}x\sqrt{g}\frac{1}{16\pi G}R-\int d^{4}x\sqrt{g}\frac{1}{16\pi \alpha_{GUT}}F^{2}.
\end{equation}
As we mentioned earlier, the values of the gauge and gravitational
couplings $\alpha_{GUT}$ and $G$ are set by the size of the extra dimensions.
This arises because the volume of the higher dimensional space in which each
particular field propagates re-scales the coupling of that field upon compactification.

For example in Heterotic M-theory, the GUT gauge field comes from the $E_8$
gauge field in the 11 D theory \cite{witten}.  This gauge field exists only in the
uncompactified directions and in the 6 extra directions which are compactified
upon a Calabi-Yau manifold.  The coupling in 4D is given by
\begin{equation}
\alpha_U=\frac{2\pi^{2/3}\kappa^{4/3}}{V}
\end{equation}
where $\kappa^{2}$ is the 11D Newton's constant and $V$ is the volume of the
Calabi-Yau.  As usual, gravity propagates in all space-time
directions including the 11th dimension and the effective
Newton's constant in 4D is given by
\begin{equation}
G=\frac{\kappa^2}{16\pi^2 V\rho}
\end{equation}
where $\rho$ is the size of the 11th (orbifolded) dimension.  So it is
possible to have variations in the 4D strength of gravity if $\rho$ evolves
over time and in the strength of both gravitational and gauge interactions if
$V$ evolves over time.  

There have been various suggestions as to the mechanisms responsible for stabilising $V$ and $\rho$ at least in weakly coupled heterotic string theory.  Perhaps the most well known are non-perturbative corrections to the scalar potential for these fields due to the condensation of the fermionic superpartners of gauge bosons \cite{volume,dilaton}, but the problem is not completely solved and becomes more complicated in the strong coupling regime.

\subsection{Other aspects of time varying extra dimensions}
In all compactifications of higher dimensional theories, if the sizes of the extra dimensions are allowed to
possess dynamics they give rise to scalar fields in the 4D theory with
cosmological equations of motion of the usual form 
\begin{equation}
\ddot{\phi}+3H\dot{\phi}+m^2\phi=0
\label{eom}
\end{equation}
where $m$ is the second derivative of the
potential in which the field is rolling and $H=\dot{a}/a$ is the Hubble expansion factor.
Variation of the size of the higher dimensions over time leads to variations
in these scalars $\phi$ and to the variation of our effective 4D gauge couplings.  
Now the observed variation of the fine structure constant is of
order $\Delta\alpha_{em}/\alpha_{em}\sim 10^{-5}$ at a redshift of order $z\sim1$
\cite{results}.  Nucleosynthesis occurs at redshifts of order $z\sim10^8-10^{10}$
and the variation in gauge couplings at that time is certainly constrained to
at least $\Delta\alpha_{em}/\alpha_{em}< 1$. If this were not the case, it would not be possible to obtain a primordial
ratio of Helium and Hydrogen that matches that observed in regions of the
universe where no nuclear processing has occurred since
nucleosynthesis \cite{camol,pagel}.  Therefore if one is to take seriously
dynamics of the form of equation ($\ref{eom}$) it is either necessary to cook up a very original potential, or to assume that the slope of
the potential is extremely small so that equation $\dot{\phi}$ is damped and
the $\phi$ is frozen at early times when
\begin{equation}
m^2\ll H^2\rm.
\end{equation}
The variation of the gauge coupling would then begin later on when the
expansion drops to the point where $H^2\sim m^2$ and the field $\phi$ is no longer frozen.  This leads to a fine tuning problem in explaining why $m$ is so small.  Also one might expect a very small mass for such a field to lead to dangerous light modes \cite{moduli} and/or have associated with it a particle leading to new long range forces \cite{dvali}.  Therefore a variation at such a late
time, if proved to be correct, would create so many problems for our existing model that it would give us a great deal of information about
not just unification but many other aspects of particle physics beyond the standard model.

\subsection{$\Lambda_{QCD}=f(\alpha_{em})$ in GUT theories}

The relationship between $\Lambda_{QCD}$ and the gauge coupling in the grand unified theory is obtained
by considering the relationship between $\Lambda_{QCD}$ and the QCD gauge
coupling $\alpha_3$ at energies just above the masses of the light quarks.  Then one can find the relation between $\alpha_3$ at these low energies and at the energy scale of unification\cite{bailin} $M_U$ to obtain 
\begin{equation}
\Lambda_{QCD}=M_{SUSY}^{\textstyle{\frac{2}{3}}}M_{U}^{\textstyle{\frac{1}{3}}}\left(\frac{m_cm_bm_t}{M_{SUSY}}\right)^{\textstyle{\frac{2}{27}}}exp\left(\frac{-6\pi}{27\alpha_3(M_{U})}\right)
\label{lamfin}
\end{equation}
where $m_i$ is the mass of the $i$ quark and we have assumed that all the SUSY
particles have the same mass which is equal to the mass of SUSY breaking
$M_{SUSY}$.  Next, by running the electroweak gauge couplings $\alpha_1$ and
$\alpha_2$ down from $M_{U}$ to low energies we obtain the relation

\begin{equation}
\left.\frac{\Delta\Lambda_{QCD}}{\Lambda_{QCD}}\right|_{M<m_c}\sim 34 \left.\frac{\Delta\alpha_{em}}{\alpha_{em}}\right|_{E\ll m_e}
\end{equation}
the linearity being applicable for variations in $\alpha_{em}$ less than a few parts in $10^3$.

It is worth pointing out that at high energies, and therefore high temperatures in the early universe, one would expect the same renormalisation to change the value of the gauge couplings and hence $\alpha_{em}$.  How do we know this is not responsible for the possible time variation of $\alpha_{em}$?  Well apart from the fact that the spectral lines correspond to a jump in electron energy and therefore are more like a tunnelling process than something like compton scattering, the variation goes the wrong way and one would expect no renormalisation of $\alpha_{em}$ below the electron mass (1 MeV, z$\sim 10^{10}$) anyway.

\section{The effect of varying gauge couplings on di-nucleon states}
\subsection{The internuclear pion potential}
If one ignores the small contribution due to the bare quark masses and the
interquark electromagnetic binding energy, the nucleon mass $M_N$ and the
masses of mesons such as the sigma and omega meson scale with $\Lambda_{QCD}$ 
\begin{equation}
M_N\propto m_{\omega}\propto m_{\sigma}\propto\Lambda_{QCD}.
\end{equation}
This is not true for the pion mass which arises as the Goldstone boson of
spontaneously broken chiral symmetry.  Goldstone bosons of exact symmetries are massless,
but the fact that the underlying chiral symmetry of the QCD Lagrangian is
already slightly broken by the presence of small but non zero quark masses
leads to a pion mass which scales as (see e.g. \cite{weinberg2})
\begin{equation}
m_{\pi}\propto (m_q\Lambda_{QCD})^{\textstyle{\frac{1}{2}}}
\end{equation}
where $m_q$ is the average mass of the light quarks.  Here we have also assumed that the energy scales of confinement and chiral symmetry breaking coincide as suggested by simulations on the lattice \cite{ourpaper}.  The exchange of pions contributes to the internuclear potential as
\begin{equation}
V(r)=-\frac{f^{2}}{4\pi}\frac{e^{-m_{\pi}r}}{r}
\label{centralpot}
\end{equation}
where the strength of the force is given by
\begin{equation}
f^2=\left(\frac{g_{\pi}^2m_{\pi}^2}{4M_N^2}\right)\propto\frac{m_q}{\Lambda_{QCD}}
\end{equation}
where $g_\pi$ is the pion nucleon coupling and $M_N$ is the nucleon mass \cite{ourpaper}.

\subsection{The stability of the di-nucleons}

The fact that the di-proton and the di-neutron are not bound is essential for the universe to evolve the way it has done.
This is because the existence of a stable di-proton would lead to
a rapid channel for the burning of hydrogen in stars. If the di-proton was
stable, it would not be possible for stars to slowly burn hydrogen on the main
sequence for long periods of time so the fact that the di-proton is unbound is
essential for our understanding of stellar evolution.  Since the internuclear potentials in the di-proton and di-neutron systems are described by the same nuclear potential (\ref{centralpot}), the only difference between the two binding criteria arises from electromagnetic effects which are very much smaller than the precision of the results in this paper.

We can use the potential (\ref{centralpot}) to calculate the amount that one
would have to change the ratio between $m_q$ and $\Lambda_{QCD}$ in order to
stabilise the di-neutron or di-proton system.  We use a trial wavefunction of the form
\begin{equation}
\psi(r)=e^{-1/m_{\pi}r}e^{-bm_{\pi}r}  
\label{wave}
\end{equation}
so that at small $r$ the second derivative of the wavefunction depends mostly upon the potential whereas at large $r$ it depends mostly upon the energy, denoted by the variational term $b$.  Then we can write   
\begin{figure}[!h]
\centering
\epsfig{file=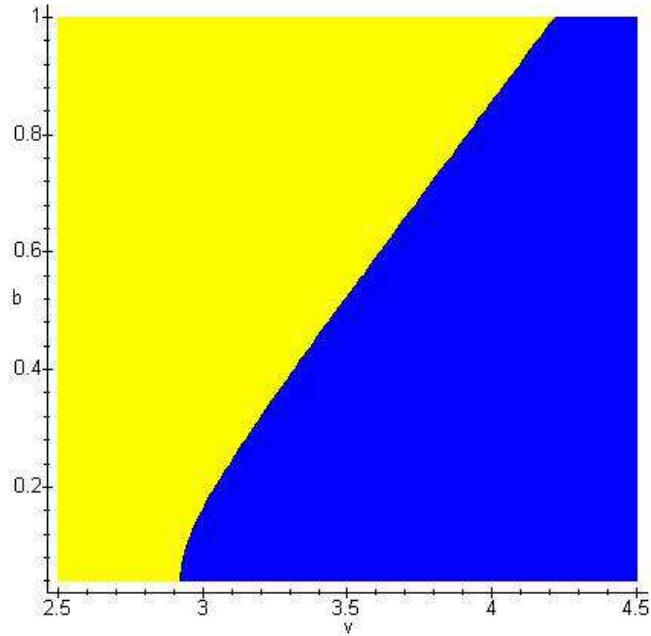,width=100mm}
\caption{Stability of di-nucleon system.  Dark region is bound, light region unbound.\label{contour}}
\end{figure}
\begin{equation}
v=\frac{f^2}{4\pi}\frac{2M_N}{m_{\pi}}\propto.
\frac{m_q}{\Lambda}
\end{equation}
Once we have evaluated $v$, we can see from figure \ref{contour} that in order for the di-nucleon
system to remain unbound,  $\Delta v/v<2.6$.  This corresponds to a very large change in $\Lambda_{QCD}$ and the corresponding change in $\alpha_{em}$ is about -5$\%$.  A similar analysis for the deuteron leads to a criterion for un-binding of about a +2$\%$ increase in $\alpha_{em}$.

\subsection{Nucleosynthesis Constraints}
We have placed constraints on the amount by which it is possible to vary
$\alpha_{em}$ so that the corresponding variation in $\Lambda_{QCD}$ is small
enough to be consistent with the stability of the deuteron and the instability
of the di-nucleon.  Of course, without saturating these bounds, a variation in
the ratio between the quark masses and $\Lambda_{QCD}$ will lead to a change
in the binding energy of the deuteron.  This will have implications for the
production of Helium during nucleosynthesis.

When the universe is hotter than about 1 MeV, neutrons and protons are
continuously interconverting via the weak interactions
\begin{eqnarray}
n+e^{+}&\leftrightarrow& p+\bar{\nu}_{e}\nonumber\\
n+\nu_{e}&\leftrightarrow& p+e^{-}\nonumber\\
n&\leftrightarrow& p+e^{-}+\bar{\nu}_{e}.\nonumber
\label{weak}
\end{eqnarray}
with the rate of these reactions being given approximately by 
\begin{equation}
\Gamma_{pe\rightarrow n\nu}\approx G_{F}^{2}T^{5}\qquad 
\label{wrate}
\end{equation}
Now when the Hubble expansion $H=1.66\sqrt{g_*G}T^2$ becomes larger than this rate the
reactions freeze out.  This occurs at a temperature
\begin{equation}
T_{o}\approx \frac{G^{1/6}}{G_{F}^{2/3}}
\end{equation}
and the ratio of neutron to proton number density $(n/p)$ which one finds at
this temperature is frozen, only changing via the beta decay of the neutron.
\begin{equation}
(n_{o}/p_{o})=e^{-(M_n-M_p)/T_{o}}
\end{equation}
Later at $T_{He}\sim B_{deut}$, the deuteron can survive without becoming
photo-dissociated by the hot plasma and the following Helium forming reactions
are able to take place 
\begin{eqnarray}
{\rm{H}}^{2}({\rm{H}}^{2},n){\rm{He}}^{3}({\rm{H}}^{2},p){\rm{He}}^{4}& & \nonumber\\
{\rm{H}}^{2}({\rm{H}}^{2},p){\rm{H}}^{3}({\rm{H}}^{2},n){\rm{He}}^{4}& & \nonumber\\
{\rm{H^{2}}}({\rm{H^{2}}},\gamma){\rm{He}}^{4}& &.\nonumber
\end{eqnarray}
Essentially all (99.99\%) of the neutrons around at this temperature form
helium via deuterium.  The mass fraction of Helium is then given by
\begin{equation}
Y=X_{4}=\left.\frac{2(n/p)}{1+(n/p)}\right|_{T_{He}}\qquad ; \qquad X_{A}=\frac{n_{A}A}{n_{N}}
\end{equation}

Now as the neutrons decay, the neutron proton ratio changes

\begin{equation}
(n/p)=\frac{exp(-t/\tau)}{(p_{o}/n_{o})+1-exp(-t/\tau)}
\end{equation}
where $\tau$ is the neutron lifetime, so the time between weak freeze out and
the temperature at which Helium can form is essential in determining the
resulting mass fraction of Helium in the universe.  Since this temperature is
determined by the binding energy of the deuteron, the process is extremely
sensitive to variations in the internuclear potential and consequently the
gauge couplings.  On finds that in order to agree with the observed primordial
mass fraction of Helium the binding energy of the deuteron is constrained to lie within
\begin{equation}
-0.20\le\frac{\Delta B_{deut}}{B_{deut}}\le +0.13
\end{equation}
which corresponds to a variation in $\Lambda_{QCD}$ of 
\begin{equation}
+0.04 \ge \frac{\Delta \Lambda_{QCD}}{\Lambda_{QCD}}\ge -0.04.
\end{equation}
If we calculate what variation in the gauge couplings can give rise to this
variation in $\Lambda_{QCD}$ we see that
\begin{equation}
-0.005\le\left.\frac{\Delta\alpha_3}{\alpha_3}\right|_{1\,{\rm TeV,\ no\ SUSY}}\le 0.005
\end{equation}
for the variation in $\alpha_3$ at 1 TeV, assuming that is where the gauge
couplings are set.  If we consider the more conventional case with SUSY and
gauge coupling unification at $10^{16}$ GeV, we can find out exactly what this
means in terms of the variation of $\alpha_{em}$
\begin{equation}
\left|\frac{\Delta\alpha_{em}}{\alpha_{em}}\right|\le 10^{-3}.
\end{equation}
This shows us that the constraint on $\alpha_{em}$ is very strong around nucleosynthesis, so if it is varying today, it cannot have varied much more rapidly in the early universe.

\section{Conclusions}

We have shown that a variation of $\alpha_{em}$ implies a much larger variation in $\Lambda_{QCD}$ if the gauge couplings are unified at very high energies.  This leads to constraints upon the variation in $\alpha_{em}$ due to the fact that we must maintain the stability of the deuteron and the instability of di-proton at all times since nucleosynthesis.  These constraints are typically of the order for a few percent.  We have also shown that it is possible to place new constraints upon the variation of $\alpha_{em}$ at nucleosynthesis since quite a small variation in $\alpha_{em}$ would give rise to a large variation in the primordial Helium abundance if one assumes gauge coupling unification. 

\section*{Acknowledgments}
This is based on work in collaboration with Thomas Dent, and I am grateful for conversations with Jean-Marie Fr\`ere.

\end{document}